\documentclass[letterpaper, 10.0 pt, conference]{ieeeconf}  
\IEEEoverridecommandlockouts                              
\usepackage{epsfig} 
\usepackage{graphicx}
\usepackage{amsmath}
\usepackage{array}
\usepackage{amssymb}
\usepackage{verbatim}
\usepackage{cleveref}
\usepackage{color}
\usepackage{relsize}
\usepackage{tikz}
\usepackage{fancyhdr}
\usepackage{graphics}
\usepackage{setspace}
\usepackage{epstopdf}
\usepackage{multirow}
\usepackage{lipsum}
\usepackage{textcomp}
\usepackage{mathtools}
\usepackage{cite}
\usepackage{setspace}
\usepackage{dblfloatfix} 
\usepackage{url}

\title{\large \bf
Predictive Second Order Sliding Control of Constrained Linear Systems\\with Application to Automotive Control Systems}

\author{Mohammad Reza Amini$^\dagger$, Mahdi Shahbakhti$^\ddagger$, and Jing Sun$^\dagger$
			\thanks{$^\dagger$~M. R. Amini and J. Sun are with the Department of Naval Architecture and Marine Engineering, University of Michigan, Ann Arbor, MI 48109 USA.	{\tt\small (\{mamini,jingsun\}@umich.edu})}%
			\thanks{$^\ddagger$~M. Shahbakhti is with the Department of Mechanical Engineering-Engineering Mechanics, Michigan Technological University, Houghton, MI 49931 USA.	({\tt\small mahdish@mtu.edu})}
}
\graphicspath{{./figures/}}

\renewcommand{\baselinestretch}{1}
\begin{document}
	\maketitle
	\thispagestyle{empty}
	\pagestyle{empty}
\begin{abstract}\\
This paper presents a new predictive second order sliding controller (PSSC) formulation for setpoint tracking of constrained linear systems. The PSSC scheme is developed by combining the concepts of model predictive control (MPC) and second order discrete sliding mode control. In order to guarantee the feasibility of the PSSC during setpoint changes, a virtual reference variable is added to the PSSC cost function to calculate the closest admissible set point. The states of the system are then driven asymptotically to this admissible setpoint by the control action of the PSSC. The performance of the proposed PSSC is evaluated for an advanced automotive engine case study, where a high fidelity physics-based model of a reactivity controlled compression ignition (RCCI) engine is utilized to serve as the virtual test-bed for the simulations. Considering the hard physical constraints on the RCCI engine states and control inputs, simultaneous tracking of engine load and optimal combustion phasing is a challenging objective to achieve. The simulation results of testing the proposed PSSC on the high fidelity RCCI model show that the developed predictive controller is able to track desired engine load and combustion phasing setpoints, with minimum steady state error, and no overshoot. Moreover, the simulation results confirm the robust tracking performance of the PSSC during transient operations, in the presence of engine cyclic variability.
\vspace{-0.15cm}
\end{abstract}

\section{INTRODUCTION} \label{Sec:Intro} 
Feasibility and stability of model predictive control (MPC) scheme for a tracking control problem, where the desired trajectory changes, are challenging to be guaranteed. Addition of a virtual reference, as an extra optimization variable, to the cost function of the MPC has been proposed in the literature~\cite{limon2008mpc,ferramosca2009mpc,ferramosca2011optimal} to solve this issue. This approach is similar to the so-called command governor technique~\cite{garone2017reference} for tracking problem of constrained linear systems, in which a nonlinear low-pass filter of the reference is added (the output of the filter can be seen as the virtual reference) to ensure the admissible evolution of the system to the reference. For the proposed approach in~\cite{limon2008mpc,ferramosca2009mpc,ferramosca2011optimal}, an additional term in the cost function of the MPC reflects the difference between the original and the virtual references. Moreover, an augmented space consisting of the original system's states and the virtual reference, is used to formulate a terminal invariant set to guarantee the persistent feasibility of the constrained optimization problem, which is then utilized to prove the asymptotic stability of model predictive controllers~\cite{ferramosca2011optimal,Andy2017}. The disadvantage of this approach is the increased complexity of the invariant set computation due to extra variable introduced to the optimization problem.

Sliding mode control (SMC) is proven to be a robust and computationally efficient solution for tracking problems of linear and nonlinear systems with a great deal of uncertainty~\cite{Slotine}. The key feature of SMC is converting a high dimensional tracking control problem into a lower dimensional stabilization control problem. On the other side, conventional SMCs do not consider the constraints on the states and inputs of the system. This is because the control input of the SMC is mainly calculated based on the present information of the system, and information from future events is not considered in the controller formulation. A novel combination of MPC and first order discrete SMC (DSMC) has been proposed in~\cite{hansen2015receding} to handle state and input constraints within the sliding controller by taking into account the future information via a receding horizon sliding control (RHSC) scheme.

The concept of second order DSMC has been proposed in the authors' previous works~\cite{Amini_DSCC2017,Amini_ACC2017}, where the asymptotic stability of the second order sliding mode controller is also proven. The main advantage of a second order DSMC in comparison with the first order controller is its enhanced smoothness in the tracking, and improved response to upcoming changes~\cite{Amini_ACC2017,Amini_DSCC2017,mihoub2009real}. This is due to the fact that in a second order DSMC, in addition to the sliding variable, its derivative is also driven to zero~\cite{salgado2004robust}. 

The second order DSMC provides fast and computationally efficient tracking performance~\cite{sira1990structure,Amini_CEP}. However, as mentioned earlier, the constraints on the states and inputs of the system can not be incorporated in the controller design, which may cause saturation in the controller system~\cite{hansen2015receding}. Therefore, formulating the second order sliding controller in a predictive scheme, not only preserves the key features of the sliding control and MPC, but also allows for handling the input and state constraints within the sliding controller formulation~\cite{mozaffari2016receding}. In this paper
, a novel \textit{predictive} second order sliding controller (PSSC) is presented. The proposed PSSC enables handling of the constraint on the inputs and states by direct inclusion of the future information. Moreover, the PSSC ensures asymptotic
tracking of any admissible piecewise setpoints, or the closest admissible setpoint, if the target trajectory is not admissible.

The performance of the proposed PSSC is evaluated for control of an advanced automotive engine including a Reactivity Controlled Compression Ignition (RCCI). RCCI technology offers significantly higher fuel conversion efficiency compared to conventional automotive engines through creating optimum heat release shape by exploiting reactivity gradients inside the combustion chamber. RCCI typically runs with lean air-fuel mixture, and has low combustion temperature, resulting in much lower \textit{NO$_x$} and \textit{PM} engine-out emissions, compared to conventional engines~\cite{splitter2010improving}. On the other side, simultaneous combustion phasing and engine load control is hard to achieve due to the strong nonlinearity and internal coupling in the dynamics of RCCI engines, specifically during the transients. It will be shown in this paper that the PSSC is able to handle the coupling within the RCCI engine to achieve simultaneous engine load and combustion phasing tracking, with consideration of the constraints on the control signals during the engine transient operation.     

The main contribution of this paper includes the first development of a predictive second order sliding controller (PSSC) for tracking of constrained linear systems. The PSSC is formulated with respect to a novel invariant sliding domain to ensure feasibility and stability of the controller. Moreover, this paper presents the first application of a predictive sliding controller for simultaneous load and combustion phasing tracking in RCCI engines. \vspace{-0.1cm}

\section{Discrete Second Order Sliding Control} \label{sec:DiscreteTimeSecondDSMC}
A linear multi-input multi-output system can be represented by the following discrete-time state space equations:
\vspace{-0.3cm}
\begin{gather}\label{eq:ACC18_Eq1}
\boldsymbol{x}(k+1)=\boldsymbol{A}\boldsymbol{x}(k)+\boldsymbol{B}\boldsymbol{u}(k), \\
\boldsymbol{y}(k)=\boldsymbol{C}\boldsymbol{x}(k), 
\end{gather}
where $\boldsymbol{x} \in \mathbb{R}^n$, $\boldsymbol{u} \in \mathbb{R}^m$, $\boldsymbol{y} \in \mathbb{R}^m$ are the state, input, and output of the linear system, respectively. $\boldsymbol{A}\in\mathbb{R}^{n\times n}$,~$\boldsymbol{B}\in\mathbb{R}^{n\times m}$,~$\boldsymbol{C}\in\mathbb{R}^{m\times n}$ are the system matrices. Moreover, it is assumed that the state and input of the linear system in Eq.~(\ref{eq:ACC18_Eq1}) are constrained: $\boldsymbol{x} \in\mathcal{{X}}\subseteq \mathbb{R}^n$, $\boldsymbol{u} \in\mathcal{{U}}\subseteq \mathbb{R}^m$.~ 
%
%
$\mathcal{{X}}$ and $\mathcal{{U}}$ sets are restricted to have the origin in their interior and be polyhedral~\cite{Andy2017}. The control objective is to drive the output of the system ($\boldsymbol{y}$) to its desired value ($\boldsymbol{y}_d$) in the presence of the state and input constraints. To this end, an error function ($\boldsymbol{e}$) is defined:
\vspace{-0.35cm}
\begin{gather}\label{eq:ACC18_Eq2}
\boldsymbol{e}(k)=\boldsymbol{y}(k)-\boldsymbol{y}_d(k)
\end{gather}
and the control objective is set to drive $\boldsymbol{e}$ to zero. 
Here, the first order sliding function ($\boldsymbol{s}$) definition of a linear system with relative degree of $d_1,\cdots,d_m$ from~\cite{Andy2017} is adopted to define the second order sliding function ($\boldsymbol{\xi}$), which will be used later to formulate the predictive sliding controller. Moreover, it is assumed that the linear system in Eq.~(\ref{eq:ACC18_Eq1}) is \textit{minimum-phase}~\cite{kraev2014generalization}. The first order sliding variable can be defined with respect to Eq.~(\ref{eq:ACC18_Eq2}) as follows:
\vspace{-0.1cm}
\begin{gather}\label{eq:ACC18_Eq3}
\boldsymbol{s}(k)=\begin{bmatrix}
s_1(k)\\
\vdots\\ 
s_m(k)\\
\end{bmatrix}=~~~~~~~~~~~~~~~~~~~~~~~~~~~~~~~~~~~~~~
\\
\begin{bmatrix}
\alpha_{_{1,0}}e_{1}(k)+\alpha_{_{1,1}}e_{1}(k+1)+\cdots+\alpha_{_{1,l_1}}e_{1}(k+l_1)\\
\vdots\\ 
\alpha_{_{m,0}}e_{m}(k)+\alpha_{_{m,1}}e_{m}(k+1)+\cdots+\alpha_{_{m,l_m}}e_{m}(k+l_m)\\
\end{bmatrix} \nonumber
\end{gather}
where $l_i=d_i-1$, and $\alpha_{i,l_i}\neq 0,~i=1,\cdots,m$. By substituting Eq.~(\ref{eq:ACC18_Eq2}) into Eq.~(\ref{eq:ACC18_Eq3}), $\boldsymbol{s}$ can be simplified as:
\vspace{-0.25cm}
\begin{gather}\label{eq:ACC18_Eq4}
\boldsymbol{s}(k)=\boldsymbol{G}\boldsymbol{x}(k)-\boldsymbol{H}(k) 
\end{gather}
where:
\vspace{-0.3cm}
\begin{gather}\label{eq:ACC18_Eq4_2}
\boldsymbol{G}=\begin{bmatrix}
\boldsymbol{c}_1\sum\limits_{j=0}^{l_1} \alpha_{_{1,j}}\boldsymbol{A}^j\\
\vdots \\
\boldsymbol{c}_m\sum\limits_{j=0}^{l_m} \alpha_{_{m,j}}\boldsymbol{A}^j\\
\end{bmatrix},\\
\boldsymbol{H}(k)=
\begin{bmatrix}
\alpha_{_{1,0}}y_{_{d,1}}(k)+\cdots+\alpha_{_{1,l_1}}y_{_{d,1}}(k+l_1)\\
\vdots\\ 
\alpha_{_{m,0}}y_{_{d,m}}(k)+\cdots+\alpha_{_{m,l_m}}y_{_{d,m}}(k+l_m) \\
\end{bmatrix}
\end{gather}
where $\boldsymbol{c}_i,~i=1,\cdots,m$ is the $i^{th}$ row of $\boldsymbol{C}$. The interesting point about Eq.~(\ref{eq:ACC18_Eq4_2}) is that $\boldsymbol{c}_i\boldsymbol{A}^{j}\boldsymbol{B}=\boldsymbol{0},~i=1,\cdots,m,~j=1,\cdots,l_i$~\cite{Andy2017}. Therefore no term containing $\boldsymbol{B}$ appears in Eq.~(\ref{eq:ACC18_Eq4_2}). If $\boldsymbol{c}_i\boldsymbol{A}^{j}\boldsymbol{B}\neq \boldsymbol{0}$, then the $i^{th}$ output of the system has a relative degree less than $d_i$, which violates the initial statement of assuming the $i^{th}$ output to have a relative degree of $d_i$, thus $\boldsymbol{c}_i\boldsymbol{A}^{j}\boldsymbol{B}= \boldsymbol{0}$.  


In the discrete time, the second order sliding variable ($\boldsymbol{\xi}(k)$) is defined with respect to Eq.~(\ref{eq:ACC18_Eq3}) as follows~\cite{Amini_ACC2017,mihoub2009real}:
\vspace{-0.25cm}
\begin{gather}\label{eq:ACC18_Eq5}
\boldsymbol{\xi}(k)=\boldsymbol{s}(k+1)+\boldsymbol{\beta}\boldsymbol{s}(k)
\end{gather}
where $\boldsymbol{\beta}\in\mathbb{R}^{m\times m}$, which manipulates the rate of state convergence to the sliding manifold, is chosen such that all the eigenvalues are inside the unit circle~\cite{Amini_DSCC2017}. The equivalent control input of the second order sliding controller is achieved by solving the following equality~\cite{Amini_ACC2017}: 
\vspace{-0.15cm}
\begin{gather}\label{eq:ACC18_Eq6}
\boldsymbol{\xi}(k+1)=\boldsymbol{\xi}(k)=\mathbf{0}.
\end{gather}
Applying Eq.~(\ref{eq:ACC18_Eq6}) leads to the following equivalent control input ($\boldsymbol{u}_{eq}(k)$) of the second order DSMC: 
\vspace{-0.15cm}
\begin{gather}\label{eq:ACC18_Eq7}
\boldsymbol{u}_{eq}(k)=~~~~~~~~~~~~~~~~~~~~~~~~~~~~~~~~~~~~~~~~~~~~~~~~~~\\
-(\boldsymbol{GB})^{-1}\Big(\big(\boldsymbol{GA}+\boldsymbol{\beta G}\big)\boldsymbol{x}(k)-\big(\boldsymbol{H}(k+1)+\boldsymbol{\beta}\boldsymbol{H}(k)\big)\Big). \nonumber
\end{gather}
Eq.~(\ref{eq:ACC18_Eq7}) calculates a vector of control input signals ($\boldsymbol{u}$) based on the model matrices, current states ($\boldsymbol{x}$), and also current and future information of the desired trajectories ($y_{d,i}(k),\cdots,y_{d,i}(k+l_i+1),~i=1,\cdots,m$). Moreover, since it is assumed that the system in Eq.~(\ref{eq:ACC18_Eq1}) has a relative degree of $d_1,\cdots,d_m$, the existence of $(\boldsymbol{GB})^{-1}$ is ensured~\cite{Andy2017,kraev2014generalization}. 

\section{Predictive Second Order Sliding Control} 
%
In this section, a novel combination of the {second order} sliding controller with MPC, as the predictive \textit{second order} sliding control (PSSC) scheme is presented for constrained linear system tracking. To this end, following the procedure proposed in~\cite{Andy2017,limon2008mpc}, first a terminal invariant second order sliding domain is calculated. Next, this invariant set is used as extended terminal constraint to formulate the PSSC. 


{In the first step, it is required to determine a terminal state feedback control law, based on the second order sliding control, in order to characterize the steady states and inputs~\cite{limon2005mpc}. Moreover, the reference output ($\boldsymbol{y}_d$) in Eq.~(\ref{eq:ACC18_Eq2}) is replaced with a virtual fixed but arbitrary reference ($\tilde{\boldsymbol{y}}_d$) to calculate the invariant second order sliding domain, according to the procedure in~\cite{Andy2017,limon2008mpc}.} By substituting $\tilde{\boldsymbol{y}}_d$ in Eq.~(\ref{eq:ACC18_Eq2}), the error becomes: $\tilde{\boldsymbol{e}}(k)={\boldsymbol{y}}(k)-\tilde{\boldsymbol{y}}_d$. It was shown in~\cite{Andy2017} that substitution of $\tilde{\boldsymbol{e}}(k)$ in Eq.~(\ref{eq:ACC18_Eq4}) results in the following first order sliding function:
\vspace{-0.15cm}
\begin{gather}\label{eq:ACC18_Eq8}
{\boldsymbol{s}}(k)=\boldsymbol{G}\boldsymbol{x}(k)-\tilde{\boldsymbol{H}}\tilde{\boldsymbol{y}}_d
\end{gather}
where:
\vspace{-0.25cm}
\begin{gather}\label{eq:ACC18_Eq9}
\tilde{\boldsymbol{H}}=\begin{bmatrix}
\sum\limits_{j=0}^{l_1} \alpha_{_{1,j}} & & \\ & \ddots & \\ & &\sum\limits_{j=0}^{l_m} \alpha_{_{m,j}} 
\end{bmatrix}
\end{gather}
Thus, the second order sliding function becomes: 
\vspace{-0.15cm}
\begin{gather}\label{eq:ACC18_Eq12}
{\boldsymbol{\xi}}(k)=\boldsymbol{G}\boldsymbol{x}(k+1)-\tilde{\boldsymbol{H}}\tilde{\boldsymbol{y}}_d+\boldsymbol{\beta}\big(\boldsymbol{G}\boldsymbol{x}(k)-\tilde{\boldsymbol{H}}\tilde{\boldsymbol{y}}_d\big), 
\end{gather}
and, the equivalent control input from Eq.~(\ref{eq:ACC18_Eq7}) is calculated as:
\small
\vspace{-0.15cm}
\begin{gather}\label{eq:ACC18_Eq10}
{\boldsymbol{u}}_{eq}(k)=
-(\boldsymbol{GB})^{-1}\Big(\big(\boldsymbol{GA}+\boldsymbol{\beta G}\big)\boldsymbol{x}(k)-\big(\boldsymbol{I}+\boldsymbol{\beta}\big)\tilde{\boldsymbol{H}}\tilde{\boldsymbol{y}}_d\Big). 
\end{gather}
\normalsize
Eq.~(\ref{eq:ACC18_Eq10}) can be written in the following form:
\vspace{-0.15cm}
\begin{gather}\label{eq:ACC18_Eq11}
{\boldsymbol{u}}_{eq}(k)=\boldsymbol{K}\boldsymbol{x}(k)+\boldsymbol{L}\tilde{\boldsymbol{y}}_d
\end{gather}
where, $\boldsymbol{K}=-(\boldsymbol{GB})^{-1}(\boldsymbol{GA}+\boldsymbol{\beta G})$ and $\boldsymbol{L}=(\boldsymbol{GB})^{-1}(\boldsymbol{I}+\boldsymbol{\beta})\tilde{\boldsymbol{H}}$. It can be seen from Eq.~(\ref{eq:ACC18_Eq11}) that the equivalent control input has two terms, the standard state feedback term ($\boldsymbol{K}\boldsymbol{x}$), and $\boldsymbol{L}\tilde{\boldsymbol{y}}_d$, which accounts for changes in the reference. 

In the second step, ${\boldsymbol{u}}_{eq}$ from Eq.~(\ref{eq:ACC18_Eq11}) is used as a terminal control input to determine the invariant set for tracking. To this end, the equivalent control input from Eq.~(\ref{eq:ACC18_Eq11}) is plugged into Eq.~(\ref{eq:ACC18_Eq1}):
\vspace{-0.15cm}
\begin{gather}\label{eq:ACC18_augmented_0}
\boldsymbol{x}(k+1)=(\boldsymbol{A}+\boldsymbol{BK})\boldsymbol{x}(k)+\boldsymbol{BL}\tilde{\boldsymbol{y}}_d.
\end{gather}
The state ($\boldsymbol{x}$) and $\tilde{\boldsymbol{y}}_d$ vectors are then augmented ($\boldsymbol{w}=[\boldsymbol{x}^{\intercal}~\tilde{\boldsymbol{y}}^{\intercal}_d]^{\intercal}$) with respect to Eq.~(\ref{eq:ACC18_Eq11}) to define the equivalent dynamic~\cite{spurgeon1992hyperplane,limon2008mpc} of the linear system in Eq.~(\ref{eq:ACC18_Eq1}):
\vspace{-0.15cm}
\begin{gather}\label{eq:ACC18_augmented}
\boldsymbol{w}(k+1)=\boldsymbol{A}_{eq}
\boldsymbol{w}(k)
\end{gather}
where:
\vspace{-0.15cm}
\begin{gather}\label{eq:ACC18_augmented_2}
\boldsymbol{A}_{eq}=\begin{bmatrix}
\boldsymbol{A+BK} & \boldsymbol{BL} \\
\boldsymbol{0}_{m\times n}     & \boldsymbol{I}_{m \times m}
\end{bmatrix}.
\end{gather}
Moreover, the state and input constraints can be re-written for the augmented system in Eq.~(\ref{eq:ACC18_augmented}) as:
\vspace{-0.15cm}
\begin{gather}\label{eq:ACC18_augmented_3}
\mathcal{W}_{eq}=\{\boldsymbol{w}:[\boldsymbol{I}_{n\times n}~~\boldsymbol{0}_{n\times m}]\boldsymbol{w}\in\mathcal{X}, [\boldsymbol{K}~~\boldsymbol{L}]\boldsymbol{w}\in\mathcal{U}\}
\end{gather}
Finally, the invariant set ($\mathcal{T}$) is calculated as:
\vspace{-0.15cm}
\begin{gather}\label{eq:ACC18_augmented_4}
\mathcal{T}=\{\boldsymbol{w}:\boldsymbol{A}_{eq}^{k}\boldsymbol{w}\in\mathcal{W}_{eq},~\forall k\geq 0\ \}
\end{gather}
and, the projected domain of $\mathcal{T}$ (Eq.~(\ref{eq:ACC18_augmented_4})) on the original state space ($\boldsymbol{x}$) is called ${\mathcal{Z}}$:
%
\vspace{-0.15cm}
\begin{gather}\label{eq:ACC18_augmented_5}
{\mathcal{Z}}=proj_{\boldsymbol{x}}(\mathcal{T}).
\end{gather}

\subsection{Predictive Controller Formulation} \vspace{-0.1cm}
In this section, the formulation of a predictive second order sliding controller (PSSC) is presented by incorporating the calculated invariant second order sliding domain for tracking from Eq.~(\ref{eq:ACC18_augmented_4}).
~The second order sliding variable from Eq.~(\ref{eq:ACC18_Eq12}) over an $N$-step prediction horizon is constructed for the PSSC, as follows:
\vspace{-0.15cm}
\begin{gather}\label{eq:ACC18_Eq18}
\boldsymbol{{\Xi}}(k+1)=[\boldsymbol{\xi}(k+1) ~ \cdots ~ \boldsymbol{\xi}(k+N)].
\end{gather}
In a similar manner, the state and input vectors over the prediction horizon are defined as:
\vspace{-0.15cm}
\begin{gather}\label{eq:ACC18_Eq19}
\boldsymbol{{X}}(k)=[\boldsymbol{x}(k)~ \cdots ~ \boldsymbol{x}(k+N)]
\end{gather}
\vspace{-0.6cm}
\begin{gather} \label{eq:ACC18_Eq19_2}
\boldsymbol{{U}}(k)=[\boldsymbol{u}(k) ~ \cdots ~ \boldsymbol{u}(k+N-1)]
\end{gather}
Eventually, based on the vectors defined in Eq.~(\ref{eq:ACC18_Eq18})-(\ref{eq:ACC18_Eq19_2}), the PSSC, as a constrained finite-time control problem, is formulated:
\vspace{-0.15cm}
\begin{gather} \label{eq:ACC18_Eq20}
\begin{aligned}
& \underset{\boldsymbol{{X}}(k),\boldsymbol{{U}}(k),\tilde{\boldsymbol{y}}_d}{\text{min}}
& & {\lVert}\boldsymbol{{\Xi}}(k+1){\rVert}_2+{\lVert}\tilde{\boldsymbol{y}}_d-\boldsymbol{y}_d(k){\rVert}_1  \\
& \text{subject to}
& & \boldsymbol{s}(i)=\boldsymbol{G}\boldsymbol{x}(i)-\tilde{\boldsymbol{H}}\boldsymbol{\tilde{y}}_d,~{\scriptstyle i=k+1,\cdots,k+N}\\
& 
& & \boldsymbol{\xi}(i)=\boldsymbol{s}(i+1)+\boldsymbol{\beta}\boldsymbol{s}(i),~{\scriptstyle i=k+1,\cdots,k+N-1} \\
& 
& & \boldsymbol{x}(i+1)=\boldsymbol{A}\boldsymbol{x}(i)+\boldsymbol{B}\boldsymbol{u}(i),~{\scriptstyle i=k,\cdots,k+N-1} \\
& 
& & \boldsymbol{x}(i)\in \mathcal{X},\boldsymbol{u}(i)\in \mathcal{U},~{\scriptstyle i=k,\cdots,k+N-1}\\
& 
& & [\boldsymbol{x}^{\intercal}(k+N)~~ \tilde{\boldsymbol{y}}_d^{\intercal}]^{\intercal}\in \mathcal{{T}}
\end{aligned}
\end{gather}
%

$\boldsymbol{\tilde{y}}_d$ in Eq.~(\ref{eq:ACC18_Eq20}) is now an optimization variable which is chosen by the PSCC optimization algorithm. Moreover, ${\lVert}\tilde{\boldsymbol{y}}_d-\boldsymbol{y}_d(k){\rVert}_1$ is an offset cost in order to account for the deviation of $\tilde{\boldsymbol{y}}_d$ from 
${\boldsymbol{y}}_d$~\cite{Andy2017,limon2008mpc}. It should be noted that the concept of second order sliding mode control is utilized to calculate the invariant set, via the equivalent control input $\boldsymbol{u}_{eq}$ from Eq.~(\ref{eq:ACC18_Eq11}), as a terminal constraint in the formulation of the PSSC. Moreover, in the absence of the constraints, and assuming a prediction horizon of $N$=1, the PSSC becomes the ideal second order DSMC. 
The feasibility and stability analysis of the proposed PSSC in Eq.~(\ref{eq:ACC18_Eq20}) can be proven by exploiting the results from~\cite{Andy2017,limon2008mpc,limon2005mpc}, with respect to the calculated invariant second order sliding domain from Eq.~(\ref{eq:ACC18_augmented_4}). 
~This analysis is skipped here, as it is out of the scope of this paper. The next section centers on illustration of the PSSC for a challenging automotive control problem. \vspace{-0.1cm}

\section{Case Study: RCCI Engine Control}
An RCCI engine, with highly nonlinear and internally coupled dynamics is chosen for the case study in this paper. The RCCI is a promising engine technology with challenging control design issues. Its performance is sensitive to operating conditions, and the combustion can become unstable if not controlled properly. Moreover, the engine, as a nonlinear dynamic plant, is subject to actuator and state constraints. These features make the RCCI engine an appropriate candidate for evaluating the proposed PSSC, as it can be designed based on a linear structure, and be implemented on the nonlinear plant while enforcing the physical constraints.\vspace{-0.15cm}

\subsection{High Fidelity Physics-Based RCCI Model} \vspace{-0.1cm}
A high fidelity physics-based RCCI engine model from~\cite{Akshat_ACC,kondipati2017modeling} is used to evaluate the performance of the proposed PSSC. The model is developed, and has been experimentally validated, to predict start of combustion (\textit{SOC}), crank angle for 50\% fuel burnt (\textit{CA50}), and indicated mean effective pressure (\textit{IMEP}).~{The details of the high fidelity physics-based RCCI dynamic model, specifications of the RCCI engine (4-cylinder 2-liter GM engine), and experimental validation results of the model are available~in~\cite{Akshat_ACC}.}

\begin{figure}[h!]
\begin{center}
\includegraphics[angle=0,width=\columnwidth]{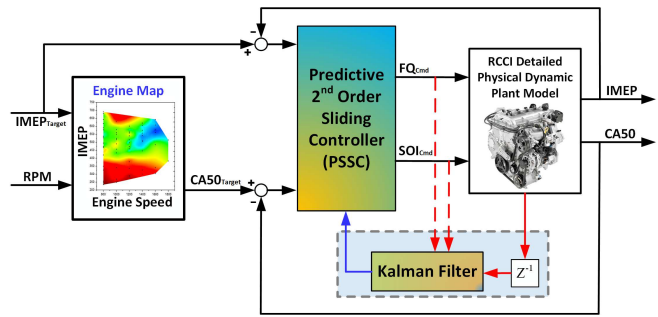} \vspace{-0.65cm}
\caption{\label{fig:PSSC_Schematic}Schematic of the designed predictive second order sliding controller for
adjusting RCCI combustion phasing (\textit{CA50}) and load (\textit{IMEP}).} \vspace{-0.75cm}
\end{center}
\end{figure}

The highly nonlinear nature of the physics-based RCCI model makes it difficult to use in the design of linear controllers. Thus, in order to design the PSSC for \textit{CA50} and \textit{IMEP} tracking, first the high fidelity model is linearized around an operation point. The linearized model has four states in the state space: \vspace{-0.15cm}
\begin{gather} \label{eq:RCCI_Linear_1}
\boldsymbol{x}=[CA50~~T_{soc}~~P_{soc}~~IMEP]^\intercal
\end{gather}
where $P_{soc}$ and $T_{soc}$ are pressure and temperature at \textit{SOC}. The output vector ($\boldsymbol{y}$) of the linear model is:  \vspace{-0.15cm}
\begin{gather} \label{eq:RCCI_Linear_2}
\boldsymbol{y}=[CA50~~IMEP]^\intercal
\end{gather}
and the control input vector is:  \vspace{-0.15cm}
\begin{gather} \label{eq:RCCI_Linear_3}
\boldsymbol{u}=[SOI~~FQ]^\intercal
\end{gather}
where, \textit{FQ} is the total injected fuel quantity, and $SOI$ is the start of injection. 
Since \textit{CA50} and \textit{IMEP} are the only measurable outputs of the RCCI model, a Kalman filter is designed to estimate the $P_{soc}$ and $T_{soc}$, with respect to the linearized model~\cite{Akshat_ACC}. The Kalman filter is updated every cycle based on the outputs of the high fidelity RCCI model. Finally, the PSSC is formulated to track the desired setpoints of \textit{IMEP} and \textit{CA50} as shown in Fig.~\ref{fig:PSSC_Schematic}. Since both outputs of the linearized model have a relative degree of one with respect to either of the inputs, the vector of the first order sliding function is defined as follows:
\begin{gather} \label{eq:RCCI_Linear_4}
\boldsymbol{s}(k)=\begin{bmatrix}
CA50(k)-CA50_{target}(k)\\
IMEP(k)-IMEP_{target}(k)
\end{bmatrix}
\end{gather}
and, the second order sliding function is calculated with respect to Eq.~(\ref{eq:ACC18_Eq5}). 
The PSSC simulations on the RCCI model are performed in MATLAB\textsuperscript{\textregistered}/SIMULINK\textsuperscript{\textregistered} by using YALMIP~\cite{lofberg2004yalmip} for formulating the optimization problem. The predication horizon is set to $N$=5 engine cycles. 
%
\vspace{-0.25cm}

\subsection{Predictive Sliding Control of the RCCI Engine}
\begin{figure}[h!]
\begin{center}
\includegraphics[angle=0,width=\columnwidth]{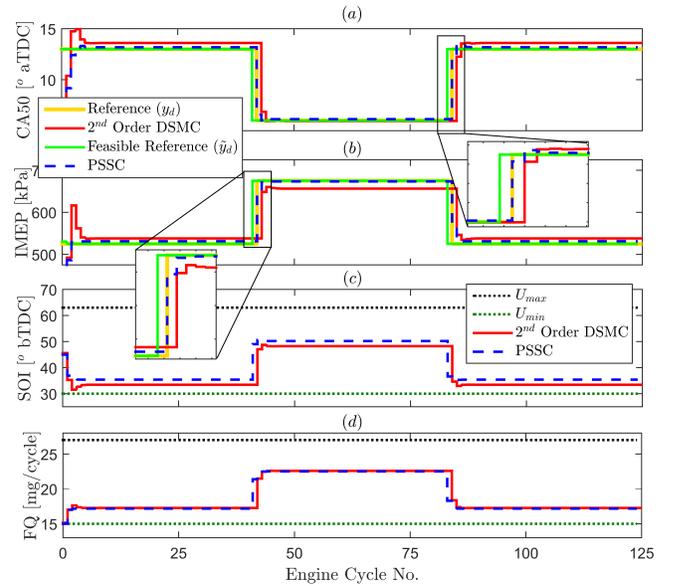} \vspace{-0.75cm}
\caption{\label{fig:TrajTrackingResults_NoDistr_PSSC_2DSMC_Real_noSat}$CA50$ and $IMEP$ control using PSSC and second order DSMC. Plant: \textbf{High Fidelity Model}. Operating conditions: $PR$= 20, $T_{in}$= 333.1 $K$, $P_{in}$= 95 \textit{kPa}, $N_e$= 1000 $RPM$, \underline{$FQ_{min}$=15 \textit{mg/cycle}}.} \vspace{-0.85cm}
\end{center}
\end{figure}

%
{Figs.~\ref{fig:TrajTrackingResults_NoDistr_PSSC_2DSMC_Real_noSat} shows the performance of the proposed PSSC and the second order DSMC in tracking variable \textit{CA50} and \textit{IMEP} trajectories, where the high fidelity physics-based RCCI model is considered as the plant according to Fig.~\ref{fig:PSSC_Schematic}. The constraints on the control inputs are selected to avoid unstable combustion. The matrix ($\mathbf{\beta}$) of the second order sliding vector is the same for the PSSC and the second order DSMC. However, there are fundamental differences between the tracking performances of these two controllers. Since the original reference ($y_d$) is feasible, the admissible reference ($\tilde{y}_d$) is the same as $y_d$. Thus, it is expected that by applying the second order DSMC to the physics-based RCCI model, the controller shows smooth and offset-free performance. However, Fig.~\ref{fig:TrajTrackingResults_NoDistr_PSSC_2DSMC_Real_noSat}-$a$ and $b$ show that overshoots occur during the early 10 cycles of the RCCI engine operation. Moreover, there are steady state errors in \textit{CA50} and \textit{IMEP} tracking, specifically at higher \textit{CA50}, and higher loads (\textit{IMEP}). This deviation in the second order DSMC performance can be explained with respect to the error in the linearization process of the highly nonlinear physics-based RCCI model around specific operation point. By comparing the second order DSMC with PSSC in Fig.~\ref{fig:TrajTrackingResults_NoDistr_PSSC_2DSMC_Real_noSat}-$a$ and $b$, it can be observed that the PSSC reacts to the upcoming changes in the reference trajectories faster than the second order DSMC. This is due to the fact that the PSSC takes into account the future information via the receding horizon strategy.}

\begin{figure}[h!]
\begin{center}
\includegraphics[angle=0,width=\columnwidth,height=7.7cm]{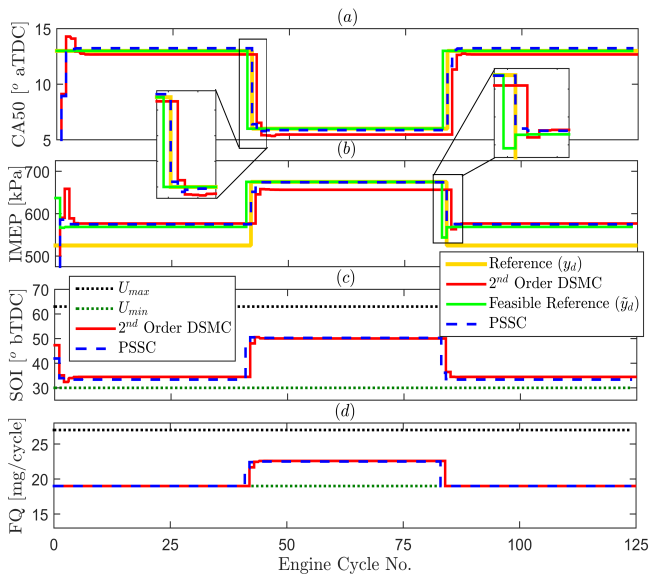} \vspace{-.85cm}
\caption{\label{fig:TrajTrackingResults_NoDistr_PSSC_2DSMC_Real_withSat}$CA50$ and $IMEP$ control using PSSC and second order DSMC. Operating conditions: $PR$= 20, $T_{in}$= 333.1 $K$, $P_{in}$= 95 \textit{kPa}, $N_e$= 1000 $RPM$, \underline{$FQ_{min}$=19 \textit{mg/cycle}}.} \vspace{-0.9cm}
\end{center}
\end{figure}

By changing the constraints on the control signals, the performance of the tracking controller is considerably affected, as the reference trajectories may not be feasible anymore. Fig.~\ref{fig:TrajTrackingResults_NoDistr_PSSC_2DSMC_Real_withSat} shows the influence of changing the operation bound of $FQ_{min}$ from 15~$mg/cycle$ in Fig.~\ref{fig:TrajTrackingResults_NoDistr_PSSC_2DSMC_Real_noSat}, to 19~$mg/cycle$. 
Due to the change in the control input constraint, the lower \textit{IMEP} level is not reachable. Since the second order DSMC does not consider the future information and cannot handle the constraints on the control input, the calculated $FQ$ signal saturates (Fig.~\ref{fig:TrajTrackingResults_NoDistr_PSSC_2DSMC_Real_withSat}-$d$), which results in a steady state error in \textit{IMEP} tracking (Fig.~\ref{fig:TrajTrackingResults_NoDistr_PSSC_2DSMC_Real_withSat}-$b$).
\begin{figure*}[htp]
\begin{center}
\includegraphics[angle=0,scale=0.48]{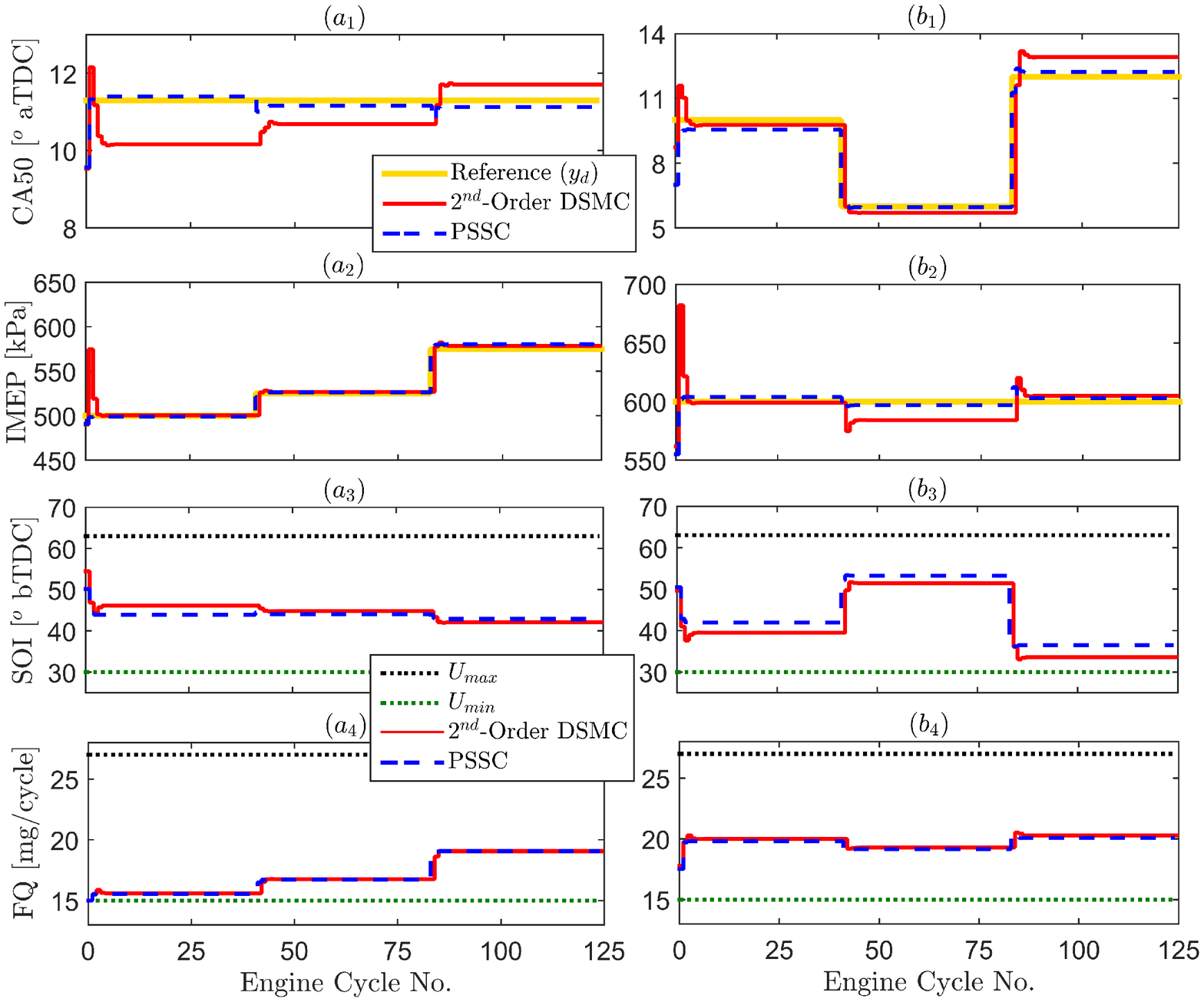} \vspace{-0.75cm}
\caption{\label{fig:TrajTrackingResults_Distrubance_PSSC_2DSMC_Real_noSat}$CA50$ and $IMEP$ control using PSSC and second order DSMC for maintaining ($a$) constant $CA50$, and ($b$) constant $IMEP$. Operating conditions: $PR$= 20, $T_{in}$= 333.1 $K$, $P_{in}$= 95 \textit{kPa}, $N_e$= 1000 $RPM$, \underline{$FQ_{min}$=15 \textit{mg/cycle}}} \vspace{-0.65cm}
\end{center}
\end{figure*}

On the other hand, the proposed PSSC calculates the feasible ($\tilde{y}_d$) with respect to actual references ($y_d$).~
Based on the calculated feasible \textit{CA50} and \textit{IMEP} references, the PSSC puts the efforts to reach the desired trajectories with lower steady state errors, and with no overshoots, compared to the second order DSMC. The better performance of the PSSC in tracking is due to its awareness of the reachable references, despite the error in the linearized RCCI model, and also the early reaction to the upcoming changes in the desired trajectories. 

The optimization algorithm of the PSSC determines the feasible reference ($\tilde{y}_d$), and provides a baseline to compare the behavior of the second order DSMC with the PSSC. 
The tracking results in Fig.~\ref{fig:TrajTrackingResults_NoDistr_PSSC_2DSMC_Real_withSat} show that while the PSSC tracks the feasible \textit{IMEP} reference accurately, the tracking performance for \textit{CA50} is not affected due to the constraint on the $FQ$, and the small steady state error is because of the error in the RCCI model linearizion. On the other side, because of the linearizion process error, and the saturation in the control signal of the second order DSMC, both \textit{CA50} and \textit{IMEP} tracking are affected, and steady state errors can be seen for both references, specifically at higher loads (\textit{IMEP}). Additionally, it can be observed that overshoots happen during the first 10 cycles of the second order DSMC operation, while the PSSC provides overshoot-free performance for both desired trajectories.

The states of the RCCI engine are highly coupled. Thus, maintaining one of the outputs at a constant level, while the other output is set to follow a variable reference is challenging to achieve. For example, it is common to maintain an optimum \textit{CA50}, while \textsc{IMEP} is changing. In order to evaluate the PSSC performance during these specific operating conditions, the simulations are performed for two different scenarios. In the first case, the objective is to maintain a constant \textit{CA50}, while the desired \textit{IMEP} is changing (Fig.~\ref{fig:TrajTrackingResults_Distrubance_PSSC_2DSMC_Real_noSat}-$a_{1-4}$). In the second case, it is desired to change \textit{CA50}, while \textit{IMEP} needs to be at a constant level of 600~$kPa$ (Fig.~\ref{fig:TrajTrackingResults_Distrubance_PSSC_2DSMC_Real_noSat}-$b_{1-4}$).  This represents situations on the RCCI engine that \textit{CA50} needs to be changed due to engine-out emission constraint, or pressure rise rate constraint.

It can be observed in Fig.~\ref{fig:TrajTrackingResults_Distrubance_PSSC_2DSMC_Real_noSat} that despite the large over shoot at the be beginning of the engine operation from the second order DSMC, both PSSC and DSMC show acceptable \textit{IMEP} tracking results (Fig.~\ref{fig:TrajTrackingResults_Distrubance_PSSC_2DSMC_Real_noSat}-$a_{2}$). However, Fig.~\ref{fig:TrajTrackingResults_Distrubance_PSSC_2DSMC_Real_noSat}-$a_{1}$ depicts that the DSMC is not able to maintain \textit{CA50} at the desired level, and there is an average error of 2~\textrm{CAD} for reference \textit{CA50} tracking from the second order DSMC. On the other side, the PSSC shows better \textit{CA50} tracking results with an average tracking error of 0.5~\textit{CAD}. 

Similarly in Fig.~\ref{fig:TrajTrackingResults_Distrubance_PSSC_2DSMC_Real_noSat}-$b$, the PSSC outperforms the second order DSMC in maintaining the desired \textit{IMEP}, while tracking the desired \textit{CA50} accurately. Specifically, it can be seen that the second order DSMC fails to keep \textit{IMEP} at 600~$kPa$ when the reference \textit{CA50} drops to 6~\textit{CAD} at the 41$^{th}$ engine cycle. Additionally, the predictive controller does not have the large overshoots and steady state error as those seen in the second order DSMC. 

\begin{figure}[h!]
\begin{center}
\includegraphics[angle=0,width=\columnwidth]{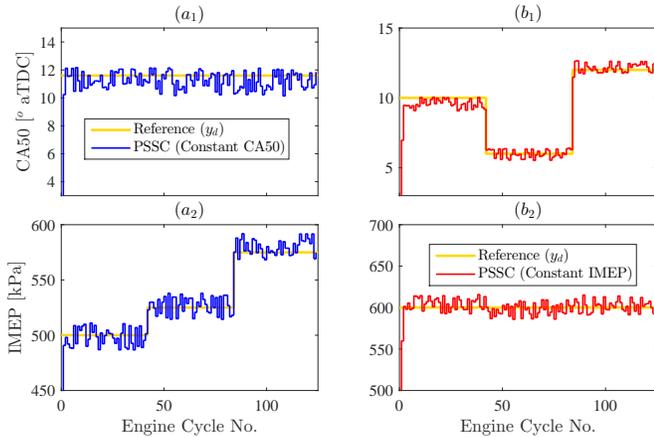} \vspace{-0.75cm}
\caption{\label{fig:CA50_IMEP_CyclicVariation_Tracking} $CA50$ and $IMEP$ control using PSSC \underline{under measurement noise and cyclic variability} for ($a$) constant $CA50$, and ($b$) constant $IMEP$. Operating conditions: $PR$= 20, $T_{in}$= 333.1 $K$, $P_{in}$= 95 $kPa$, $N_e$=1000 $RPM$, cyclic variability of 2$^o$ for $CA50$, and 25~$kPa$ for $IMEP$, based on the RCCI engine experimental data in~\cite{Akshat_ACC}.}  \vspace{-0.8cm}
\end{center}
\end{figure}

Finally, in order to evaluate the PSSC under the RCCI engine cyclic variability and measurement noise, the PSCC is studied in 
Fig.~\ref{fig:CA50_IMEP_CyclicVariation_Tracking} for tracking the same \textit{CA50} and \textit{IMEP} trajectories evaluated in Fig.~\ref{fig:TrajTrackingResults_Distrubance_PSSC_2DSMC_Real_noSat}. 
It can be observed that under the introduced measurement noise and cyclic variability, the PSSC is able to achieve the desired loads (\textit{IMEP}), while maintaining \textit{CA50} at a constant level (Fig.~\ref{fig:CA50_IMEP_CyclicVariation_Tracking}-$a_{1,2}$). Similarly, Fig.~\ref{fig:CA50_IMEP_CyclicVariation_Tracking}-$b_{1,2}$ illustrates good PSSC performance with average tracking errors of 11.4 \textit{kPa} and 0.4 \textrm{CAD} in maintaining the engine at a constant \textit{IMEP}, while desired \textit{CA50} changes. 

%

\vspace{-0.05cm}
\section{Summary and Conclusions}   \label{sec:Conclusion} \vspace{-0.1cm}
A new formulation of a predictive sliding controller based on the concept of second order sliding mode, in combination with MPC, was presented in this paper for tracking of constrained linear systems. A virtual reference, as the admissible reference, was added to the cost function of the predictive controller to account for the changes in the setpoint. Since inclusion of the virtual reference requires computation of a terminal set to guarantee the feasibility of the proposed predictive second order sliding controller (PSSC), an extended invariant second order sliding domain was calculated based on the augmented state space of the system states and the virtual reference. The performance of the proposed PSSC was demonstrated in a multi-input multi-output structure for a highly nonlinear and internally coupled RCCI engine tracking problem. An experimentally validated physics-based model of an RCCI engine was used to design and assess the proposed PSSC for simultaneous load (\textit{IMEP}) and combustion phasing (\textit{CA50}) tracking under hard constraints on the control inputs of the engine. Comparing to the second order DSMC, the proposed PSSC showed better tracking results, with minimum steady state error, and no overshoot for different tracking scenarios. Moreover, the simulation results confirm the robustness of the PSSC under measurement noise and the engine cyclic variability. \vspace{-0.15cm} 


\small
\section*{ACKNOWLEDGMENT}
Dr. Andreas Hansen from University of California, Berkeley, and Mr. Akshat Raut from Michigan Tech are gratefully acknowledged for their technical comments during the course of this study.\vspace{-0.3cm}
\bibliographystyle{unsrt} 
\bibliography{ACC2017Ref.bib} \vspace{-0.25cm}
%

\end{document}